\documentstyle[12pt,psfig]{article}
\begin{document}
\setlength{\topmargin}{-1cm}
 \setlength{\baselineskip}{18pt}
 \setlength{\textheight}{24cm}
 \setlength{\textwidth}{22cm}

\title{Critical Analysis of Electronic Simulation of Financial
Market Fluctuations}

\author{H. Fanchiotti, C.A. Garc\'{\i}a Canal
 and N. Mart\'{\i}nez \footnote{E-mails:
huner@fisica.unlp.edu.ar; garcia@fisica.unlp.edu.ar; nolo@fisica.unlp.edu.ar} \\
Departamento de F\'{\i}sica, Universidad Nacional de La Plata \\ C.C.
67 - 1900 La Plata, Argentina.}
\date{\today}
\maketitle
\begin{center}
{\bf Abstract}\end{center} {An interesting analog circuit for
simulating a signal with fluctuations having a probability
density function with a power tail has recently been proposed and
constructed. The exponent of the power law can be fixed by tuning
an appropriate circuit element. The proposal is to use the
circuit as a simulator-generator of financial market fluctuations
and as a tool for risk estimations and forecasts. We present a
discussion of the stability conditions for multiplicative noise
and an exhaustive analysis of the power law fluctuation generator
in
        connection with the electronic components and their parameters.
From our studies one can conclude that the proposal is not
adequate to provide a confident experimental scheme to follow the
fluctuations of the financial market due to both electronic
implementation and to difficulties in parameter tuning.}

\section{Introduction}
Recently, Sato, Takayasu and Sawada \cite{jap} have constructed a
very interesting analog circuit able to generate a signal
corresponding to a time series with fluctuations resembling those
of a typical financial market. The corresponding probability
density function presents a power law tail.

It is worth mentioning that the proposal of Ref. \cite{jap}
immediately attracted the opinion of the scientific community. In
particular, we should mention that the American Institute of
Physics Bulletin of Physics News number 478 of April 6, 2000
includes a comment on Ref. \cite{jap} under the title "An
electrical circuit mimics yen-dollar fluctuations". It ends by
saying that the circuit costing approximately $\$5$ can estimate
yen-dollar fluctuations as fast as the $\$10,000$ workstations
that are running mathematical simulations of the exchange rates.
On the other hand, CERN Courier of June 2000 includes also a
comment under the title "How can you make a million?" where one
can read that Japanese Scientists have unveiled a $\$5$
electrical circuit that can mimic fluctuations in the yen-dollar
exchange rate.

The circuit constructed in Ref. \cite{jap} being very interesting
in its own, deserves a careful analysis from the electronic point
of view. We present here an exhaustive study of the circuit
behavior and of its potentialities for reproducing financial
market fluctuations. We conclude that the circuit really mimics
these fluctuations but the meaning of this mimics is precisely
the one the Oxford Dictionary gives: imitate closely in order to
make fun of.

Before the presentation of the circuit and its relationship with
the Langevin equation with multiplicative noise, we discuss, in
Section 2 the stability conditions for this kind of noise.
Afterwards we include in Section 3 an analysis of the power law
fluctuations generator in connections with the electronic
components and their corresponding parameters. A circuit
simulation completes our study and its results are presented in
Section 4. Finally, Section 5 summarizes our conclusions.

\section{Stability}

Before entering into the discussion of both the electronic and
the simulation analysis, let us briefly review the stability in
multiplicative stochastic processes. It is well known that an
additive stochastic process is stable whenever the associated
deterministic problem is. In fact, given the process defined by
\begin{equation}
\frac{dv}{dt} = L(v) + \eta
\end{equation}
where $\eta$ is a standard Gaussian noise, it has a stable
stationary behaviour when all its moments $\left< v^n \right>$
exist up to the n-th order. This imposes a condition on the
integral of $L(v)$ \cite{SB}.

 However, for a multiplicative noise the
stability of the deterministic problem is not enough to ensure the
stability under fluctuations \cite{SB}. Consider now the process
\begin{equation} \label{lm}
\frac{dv}{dt}= -d\,v + v\,\eta
\end{equation}
where $\eta$ is again a standard Gaussian noise and $d$ a
parameter. In this case the moments are given by
\begin{equation}\label{vn}
\left< v^n \right> = \left< v^n \right>_{t=0}\,
exp\left[-n\,t\,\left(d -n\,D/2\right)\right]
\end{equation}
where $D$ is the measure of the Gaussian fluctuation, namely
\[
\left<\eta(t+\tau)\,\eta(t) \right> = D\,\delta(\tau)
\]
The stationary distribution results
\begin{equation}
p_0(v) = v^{-1-2\,d/D}
\end{equation}
which is not normalizable. It is then clear from Eq. (\ref{vn})
that in this case of multiplicative noise, the condition
\[
n\leq\frac{2\,d}{D}
\]
has to be fulfilled in order to guarantee the finiteness of the
moments $\left< v^n \right>$ when $t \rightarrow \infty$.
Consequently, if $d < D/2$, the multiplicative noise overcome the
deterministic restoring term $-d\,v$, even if this deterministic
problem has a stable solution.

Having these mathematical conditions in mind, we re-analyzed the
circuit proposed in Ref \cite{jap}. The main point is to
simulate, electronically or numerically, the equation of Langevin
type with multiplicative noise (\ref{lm}), that in the present
case reads
\begin{equation}
\frac{d v_0}{d t} = \left(-\frac{1}{R_f\,C_f} +
\frac{k}{R_v\,C_f}\,\mu(t)\right)\,v_0 + \psi(t)  \label{EJ}
\end{equation}
Notice a change of sign in the first term of the r.h.s of this
equation, with respect to Eq. (12) of Ref.\cite{jap}.

\section{Electronics}

\subsection{Generalities}

\begin{figure}[htbp]
\vspace*{13pt}
\centerline{\psfig{file=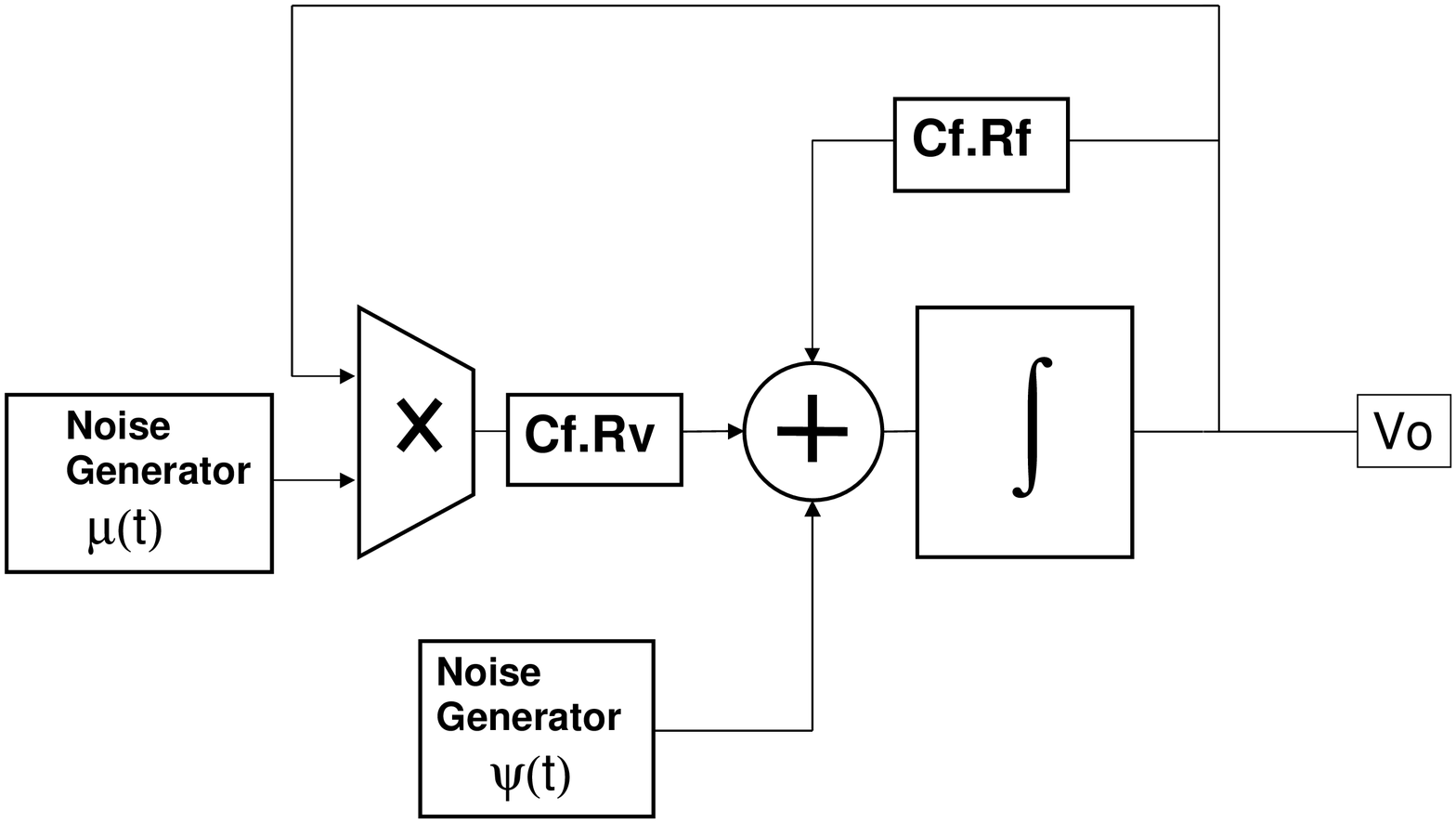,height=5cm,width=14cm}}
\caption{Block diagram of analog simulation} \label{fig:DB}
\end{figure}

The general block diagram in Fig. \ref{fig:DB} represents the
integrated form of the original differential equation \ref{EJ}.

\begin{figure}[htbp]
\vspace*{13pt} \centerline{\psfig{file=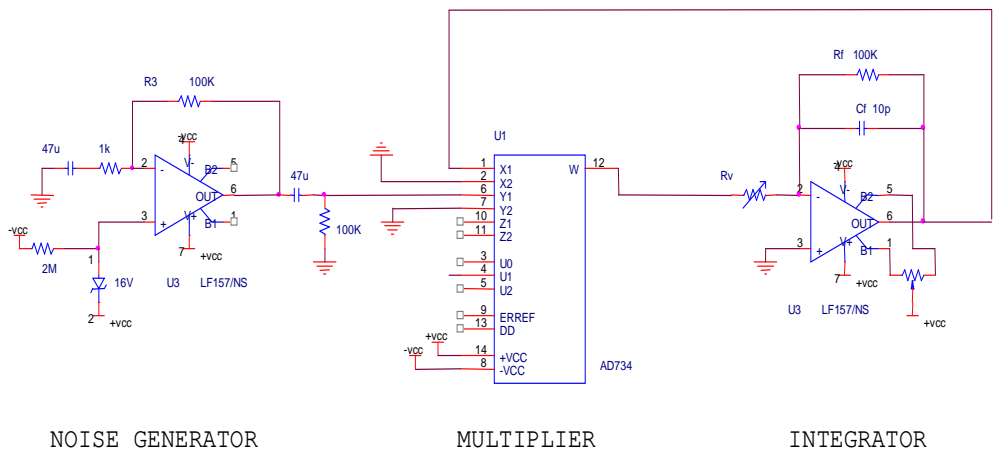,height=8
cm,width=14cm}} \caption{Power fluctuation generator}
\label{fig:PFG}
\end{figure}

The different blocks involved in Fig. \ref{fig:DB} are
implemented by using electronic components, mainly amplifiers.
From the diagram in Fig. \ref{fig:DB}, the electronic
implementation can be easily derived giving rise to the circuit
in Fig. \ref{fig:PFG}

The electronic components have their own characteristics and
impose some restrictions on the functions to be implemented. It
means that the range of values where the function is valid is
bounded by the electronic components behaviour.

\begin{figure}[htbp]
\vspace*{13pt}
\centerline{\psfig{file=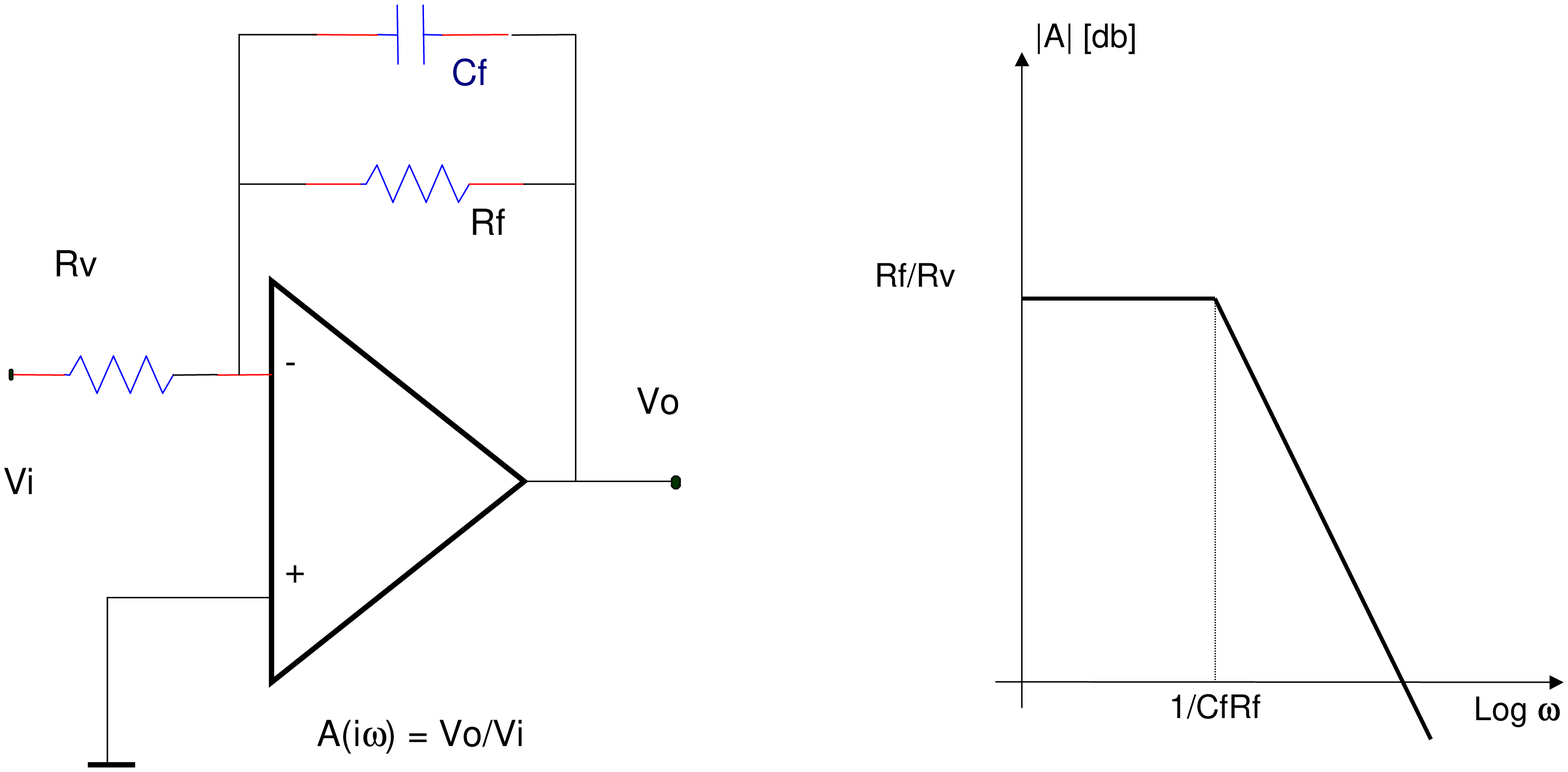,height=7cm,width=14cm}}
\caption{Electronic integrator and its Bode plot} \label{fig:EI}
\end{figure}

Let us then analyze, following Ref. \cite{HH}, the integrator
block implementation shown in Fig. \ref{fig:EI}.

Considering an ideal amplifier, namely one with Gain Product
Bandwith (GBW)$\rightarrow \infty$ and Gain $(A_0)\rightarrow
\infty$, the corresponding transfer function is
\begin{equation} \label{tf}
A(s) = \frac{V_0(s)}{V_i(s)} =- \frac{R_f}{R_v}\,\frac{1}{\left(1
+ s\,R_f\,C_f\right)}
\end{equation}
 with $s = i\,\omega$.
 Fig. \ref{fig:EI} includes also the Bode plot of this expression.

Under the condition
 $\omega \gg 1/(C_f\,R_f)$, Eq (\ref{tf}) gives rise to
 \begin{equation}
V_0(s) = -\frac{R_f}{R_v}\,\frac{1}{\left(
s\,R_f\,C_f\right)}\,V_i(s)
\end{equation}
and
\begin{equation}
v_0(t) = - \frac{1}{R_f\,C_f}\,\int v_i(t)\,dt
\end{equation}
Consequently the above circuit works as an integrator for
frequencies above $\omega_p =1/(R_f\,C_f)$. In summary, to
implement an integrator using a real operational amplifier (OA)
its GBW must be larger than the highest frequency of interest and
$A_0 \gg R_f/R_v$.

To verify the above conditions, the Bode plot of the desired
integrator, should be compared with that of the open loop gain of
the OA. As a rule of thumb, the implementation will be valid if
the plot of the OA "contains" the integrator's Bode plot.

\subsection{Circuit analysis}

The criteria above was applied to the original circuit proposed in
Ref. \cite{jap} that includes the values
\[
C_f = 10\,pF\,\,\,;\,\,\,R_f = 100\,K\Omega\,\,\,;\,\,\,R_v = 5 -
200\,\Omega
\]
giving rise to
\[
\omega_p= 1\, MHz \,\,\,;\,\,\,  A= \frac{R_f}{R_v}= 54 - 86\, db.
\]

\begin{figure}[htbp]
\vspace*{13pt}
\centerline{\psfig{file=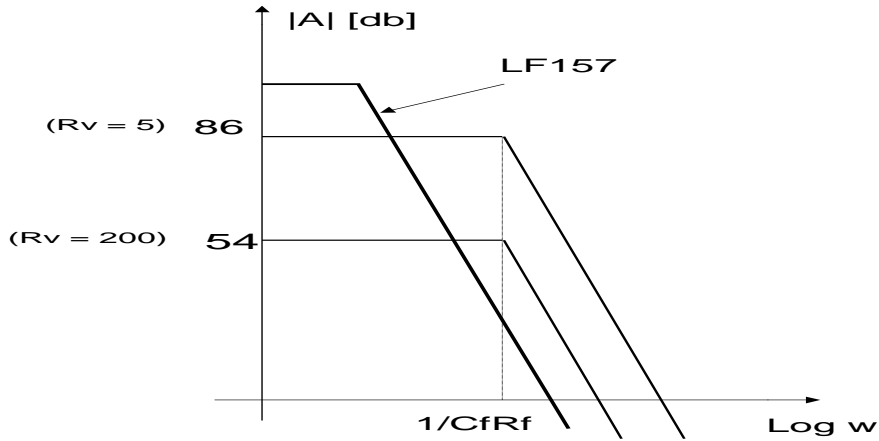,height=9cm,width=14cm}}
\caption{Comparison between Bode plots for the amplifier and the
integrator} \label{fig:BD}
\end{figure}

In Fig. \ref{fig:BD}, a Bode diagram of Eq. (\ref{tf}) is plotted
together with the OA $LF157$ characteristic obtained from the
National Semiconductor Co. Linear Databook. As can easily be
seen, the condition that the OA "contains" the desired
characteristic is not fulfilled. So the changes of $R_v$ between
the specified values do not produce a full change in the
''integrator gain'' and the observed effects are only marginal
ones. In other words, a change in $R_v$ is not fully reflected as
a change of the ''integrator gain''.

\begin{figure}[htbp]
\vspace*{13pt}
\centerline{\psfig{file=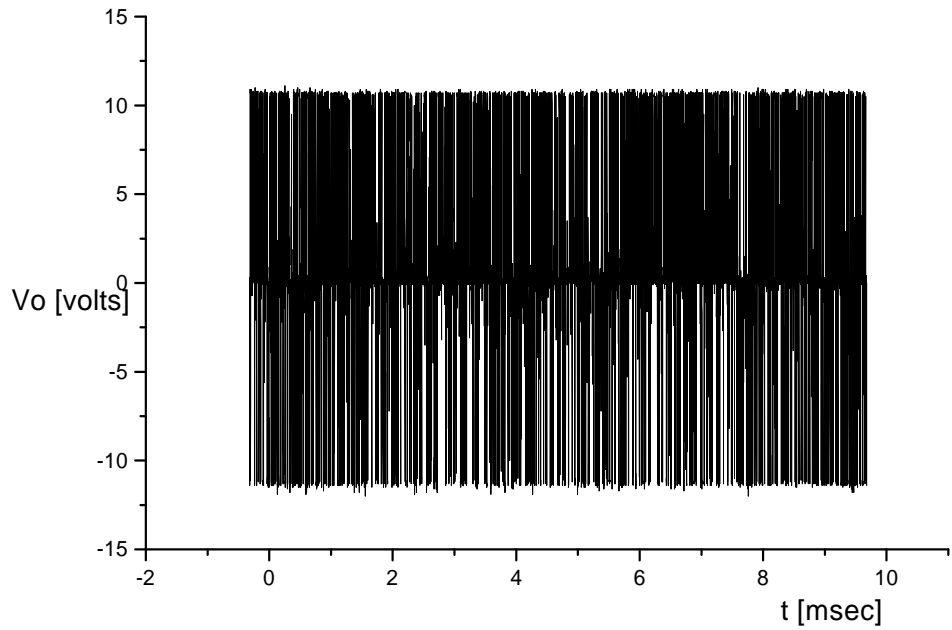,height=10cm,width=14cm}}
\caption{Output of the circuit with the original parameters for
$R_v=35\,\Omega$ } \label{fig:OJa}
\end{figure}

Moreover, when the circuit in Fig. \ref{fig:PFG} was tested with
the specified values of $R_v$, the output presents unwanted
saturation as shown in Fig. \ref{fig:OJa}.

\begin{figure}[htbp]
\vspace*{13pt}
\centerline{\psfig{file=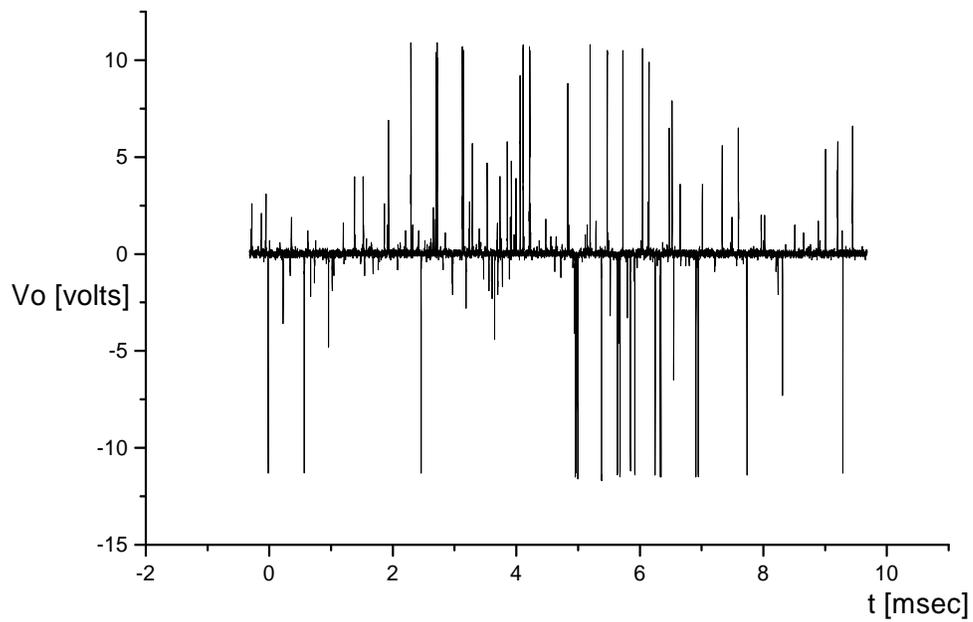,height=10cm,width=14cm}}
\caption{Output of the circuit with the original parameters for
$R_v= 2200\,\Omega$} \label{fig:OJb}
\end{figure}

Only for values of $R_v > 1.7\, K\Omega$, the amplifier remains
almost linear giving rise to the output seen in Fig.
\ref{fig:OJb}. In fact, the reported behaviour in Ref. \cite{jap}
was qualitative found for $1.7\,K\Omega <R_v< 4\,K\Omega$. For
larger values, the electrical noise dominates.

It is important to remark that our measurements were done with a
Tek scope (Model TDS 3030) with 8 bits of resolution.
Nevertheless we are certain that this technical detail doesn't
change our conclusions.

Following the circuit criteria discussed above, a new set of
values for $R_v$, $C_f$ and $R_f$ was chosen . To include the
integrator characteristic inside the Operational Amplifier (OA)
open loop, a lower pole frequency ($\omega_p$) was in order. To
avoid output saturation, higher values of $R_v$ are also needed.
Then, we decided to use the values
\[
C_f= 11\, pF\,\,\,;\,\,\, R_f= 800\, K \Omega \,\,\,;\,\,\, R_v =
6 - 12 K\Omega.
\]
  Resulting in
\[
\omega _p= 113\, KHz\,\,\,;\,\,\, A= \frac{R_f}{R_v} = 35-45\, db
\]
This is not an arbitrary selection because a value of $C_f$
higher than $15\, pF$ put the system in a heavy oscillation. A
careful study of the LF157 characteristics shows that the
stability conditions are greatly affected if a capacitor larger
than $15\, pF$ is connected between the input and the output.
Consequently, a different amplifier should be used instead.

\begin{figure}[htbp]
\vspace*{13pt}
\centerline{\psfig{file=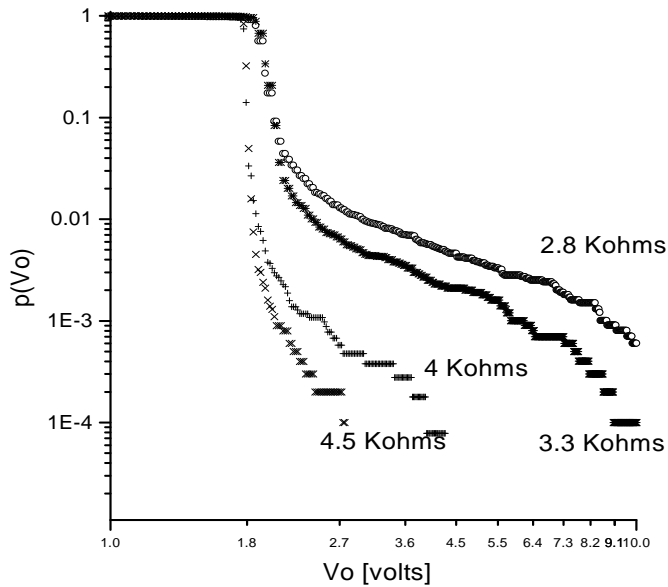,height=9cm,width=14cm}}
\caption{Probability density function for the modified
parameters} \label{fig:FB}
\end{figure}

The circuit with modified parameters worked as predicted. In Fig.
\ref{fig:FB} we present the log-log plot of the probability
density function for the $R_v$ values specified.

Using the adequate values for $C_f\,R_f$, a tighter control of the
integrator gain is obtained. Notice also that for lower values of
$R_v$, that means higher gain, the circuit does not met the
integrator condition, and for $R_v$ higher than $12\, K\Omega$
the electrical noise dominates.

\section{Simulation}

In order to check the influence of the amplifier's offset voltage
we have performed additional tests. This offset, independently of
the circuit values, does not affect the final power law behaviour
and can be reduced in the data processing stage.

If the circuit is to be really useful in predicting fluctuations
on a quantitative way, it must be necessary to know the influence
of each electrical parameter on the output signal in order to
reproduce, for example, the probability distribution of
historical data. This work would be tedious, when not impossible
using the real circuit.

We propose to use instead some simulation procedures that, without
loosing the electrical point of view, put in evidence the
influence of each of the parameters. We have tried several
configurations using the Spice program \cite{spice}. Although
this program allows one to use realistic components, by means of
transistor- level models, we prefer to use quite ideal
representation where the different parameters of the circuit,
desirable or not, can be added separately. It can be mentioned
that the simulation with real components simulation is also
plagued of serious convergence problems, being certainly
difficult to make it runs in a fiable way.

\begin{figure}[htbp]
\vspace*{13pt}
\centerline{\psfig{file=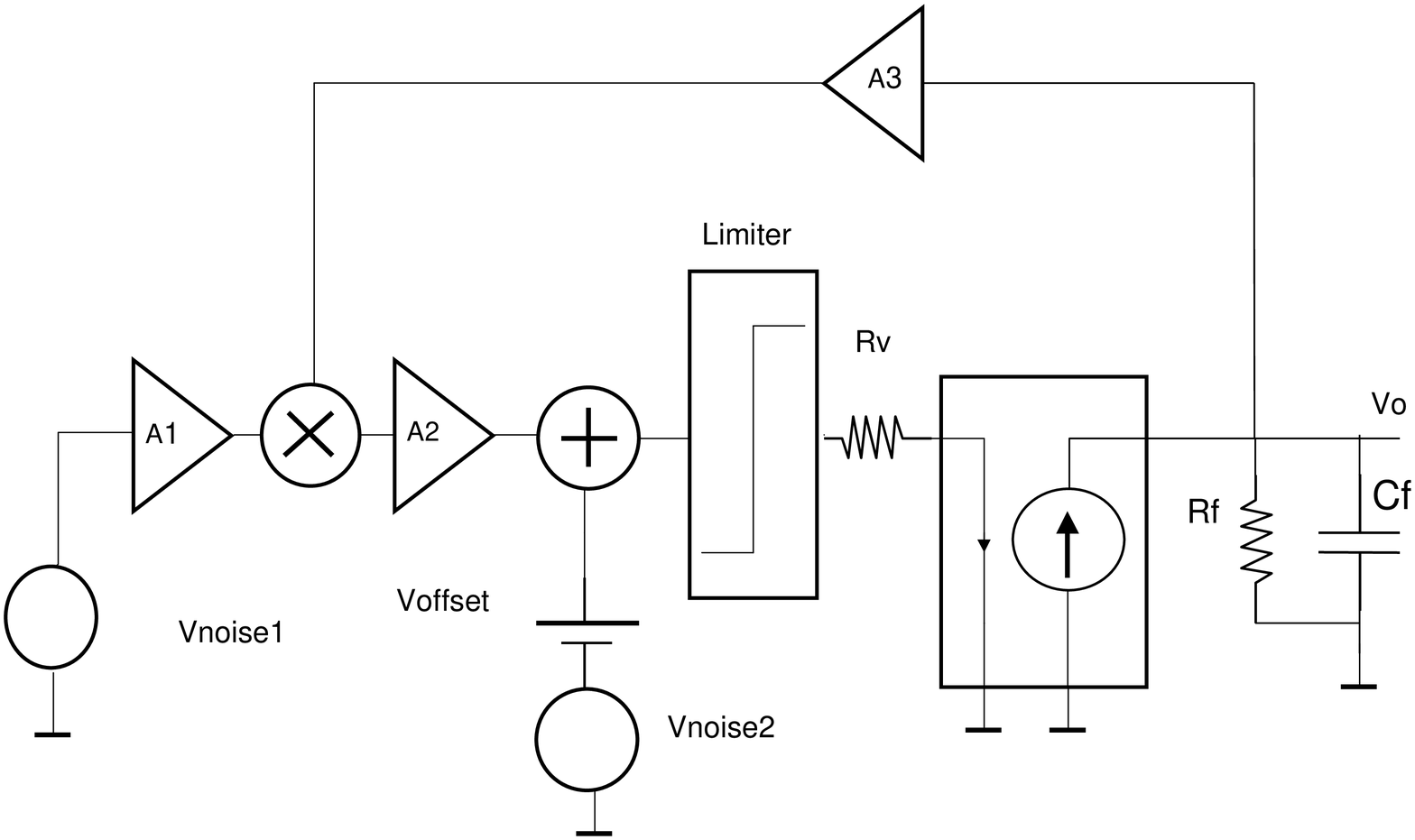,height=10cm,width=14cm}}
\caption{Simulated circuit} \label{fig:CS}
\end{figure}

Fig. \ref{fig:CS} shows the circuit used in the simulation. It
reproduces the differential equation of interest. Different gain
blocks allows one to change amplitudes, parameters such as offset
voltage and saturation limits when added, as well as different
noise generators.

\begin{figure}[htbp]
\vspace*{13pt}
\centerline{\psfig{file=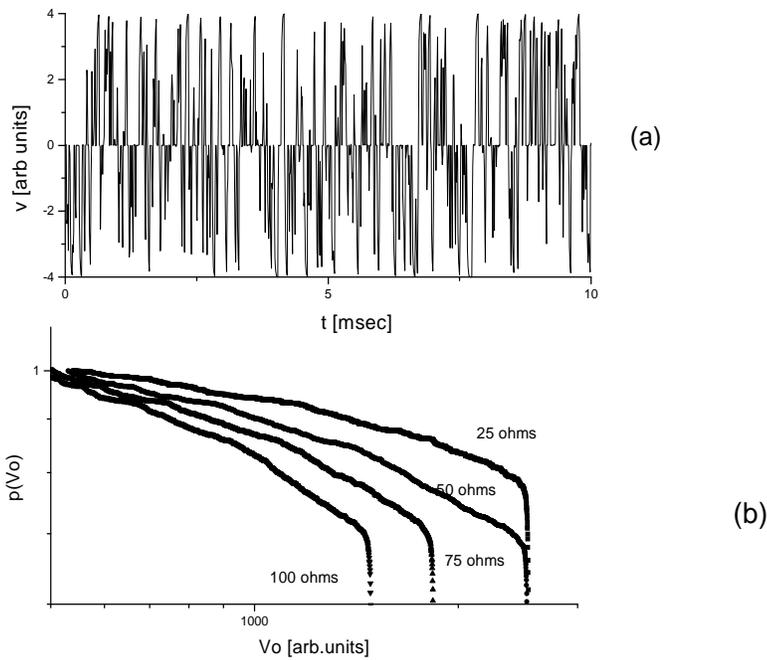,height=10cm,width=14cm}}
\caption{Output of the simulation. a)signal; b) density of
probability} \label{fig:RS}
\end{figure}

The typical output of the simulation is summarized in Fig.
\ref{fig:RS}.

\section{Conclusions}

We have carefully analyzed the recently proposed electronic analog
circuit that generates fluctuations with a probability density
function having power law tails.

We found that the originally proposed parameters are not the most
pertinent ones because they place the circuit in the borderline
of the integrator Bode plot. With appropriate parameters, the
circuit performance is noteworthily improved. Moreover, by
changing the integrated circuit originally chosen, a further
improvement can be achieved.

On the other hand, the numerical simulation, even if carried out
in terms of ideal electronic components, reinforces the previous
conclusions.

Finally, several comments are in order:
\begin{itemize}
\item Even if the power fluctuation generator circuit is certainly
a low cost apparatus, a costly peripheral equipment is needed for
taking  and  analyzing the output data.
\item The circuit, in order to mimic reasonable fluctuations, needs a fine
tuning of the pertinent parameters.
\item A direct numerical simulation of the original equation seems
to be a much better solution.
\end{itemize}

\section{Acknowledgment}
We want to warmly thank A.-H. Sato for correspondence regarding
reference \cite{jap}.

%\end{document}

\end{document}